\documentclass{PoS}

\usepackage{graphicx,epsfig,epstopdf}
\usepackage{bm,bbm}
\usepackage{braket}
\usepackage{slashed}
\usepackage{multirow}
\usepackage{float}
\usepackage{amsmath,amssymb}

\newcommand{\be}{\begin{equation}}
\newcommand{\ee}{\end{equation}}
\newcommand{\bea}{\begin{eqnarray}}
\newcommand{\eea}{\end{eqnarray}}

\def\lQ{\Lambda_{\rm QCD}}
  
\def\als{\alpha_{\rm s}}

\title{Quarkonium hybrids}

\ShortTitle{Quarkonium hybrids}

\author{\speaker{Antonio Vairo}  \\
        Physik Department, Technische Universit\"at M\"unchen, D-85748 Garching, Germany \\
        E-mail: \email{antonio.vairo@tum.de}}

\abstract{We discuss the quarkonium hybrid spectrum from the perspective of nonrelativistic effective field theories combined with lattice QCD.}

\FullConference{The European Physical Society Conference on High Energy Physics\\
                 5-12 July\\
                 Venice, Italy}

\begin{document}

\section{Quarkonium hybrids and the nonrelativistic expansion}
Among the many new quarkonium states discovered at the B factories some may be hybrids.
A {\em quarkonium hybrid} consists of a heavy quark-antiquark pair, $Q\bar{Q}$, in a color octet configuration bound with gluons.
Many approaches have been developed over the years to describe hybrids:
constituent gluon picture, Born--Oppenheimer approximation, flux tube model ...
In the following, we will argue that combining nonrelativistic effective field theories and lattice QCD 
provides a method to study systematically hybrids made of heavy quarks in QCD. 

A systematic approach based on nonrelativistic effective field theories to describe also quarkonium excitations
due to gluons has been set up in~\cite{Brambilla:1999xf,Brambilla:2000gk,Pineda:2000sz}.
An extension to include excitations due to light quarks, i.e., tetraquarks, has been suggested in~\cite{Brambilla:2008zz,Braaten:2013boa},
and explicitly worked out in~\cite{Braaten:2013boa,Braaten:2014ita,Braaten:2014qka}.
An effective field theory that incorporates the Born--Oppenheimer approximation has been proposed in~\cite{Berwein:2015vca,Brambilla:2017uyf}.
A similar approach is in~\cite{Oncala:2017hop,Soto:2017one}.

The reason why quarkonium hybrids are nonrelativistic systems is that the heavy quark mass $M$
is the largest scale in these systems, i.e., larger than the typical relative momentum $p$ of the heavy quarks
and larger than the hadronic scale $\lQ$.
The inequality $M \gg p$ implies that the systems are characterized by the hierarchy of scales typical of a nonrelativistic bound state: $M \gg p \gg E$,
where $E$ is the typical energy of the system.
Systematic expansions in the ratios of the scales $p/M$, $E/p$ and $E/M$ may be implemented at the
Lagrangian level by constructing suitable {\em nonrelativistic effective field theories}~\cite{Brambilla:2004jw}.
The ultimate effective field theory is {\em potential nonrelativistic QCD} (pNRQCD),
whose degrees of freedom  are $Q\bar{Q}$ color singlet and color octet fields, low-energy gluons and light quarks.
Note that the hierarchy of nonrelativistic scales makes the very difference of quarkonium-like systems from heavy-light mesons,
which are characterized by just two scales: $M$ and $\lQ$.
The inequality $M \gg \lQ$ implies $\als(M) \ll 1$: phenomena happening at the scale $M$ may be treated perturbatively.
Nevertheless, we may further have small couplings if, for instance, $p \gg \lQ$, in which case  $\als(p) \ll 1$.
This is likely to happen only for the lowest states. 

Consistently with the picture of adiabatically moving heavy quarks,
quarkonium hybrids are described in pNRQCD as heavy quark-antiquark pairs 
moving in the potentials $E_\Gamma$ generated by the gluons~\cite{Brambilla:1999xf,Berwein:2015vca,Brambilla:2017uyf}.
At leading order in the nonrelativistic expansion, $E_\Gamma$ is given by the energy levels of a static quark and antiquark pair.
A large number of states can be built out of the vibrational modes of each potential.

The energies $E_\Gamma$ can be computed from suitable matrix elements by means of lattice QCD.
The hybrid energy levels are obtained by solving at the level of pNRQCD the Schr\"odinger equation with the Hamiltonian $H_{\rm kin} + E_\Gamma$,
where $H_{\rm kin}$ stands for the hybrid kinetic energy operator.
$H_{\rm kin}$ acts also on the gluonic components of the wave functions and can mix different gluonic excitations.
Mixing is specially relevant for gluonic excitations that are not well separated.

\begin{figure}[ht]
\makebox[0cm]{\phantom b}\put(110,0){\epsfxsize=8truecm \epsffile{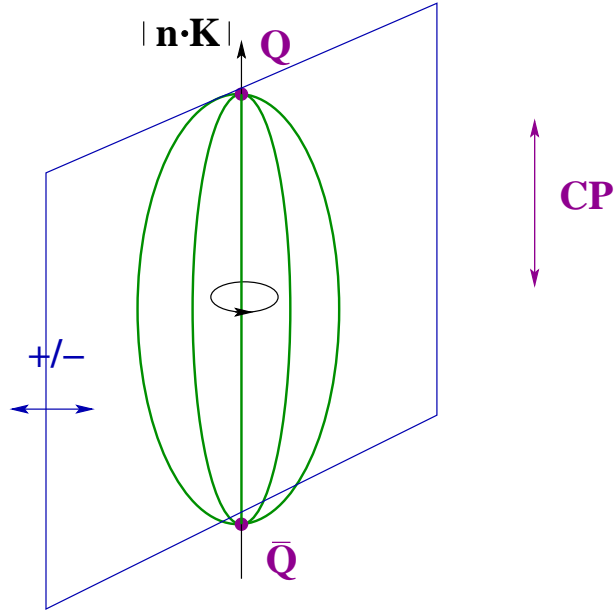}}
\caption{Quarkonium hybrid symmetries.\label{figsym}}
\end{figure}

\section{Symmetries}
\begin{figure}[ht]
  \makebox[0cm]{\phantom b}\put(110,0){\epsfxsize=8truecm \epsffile{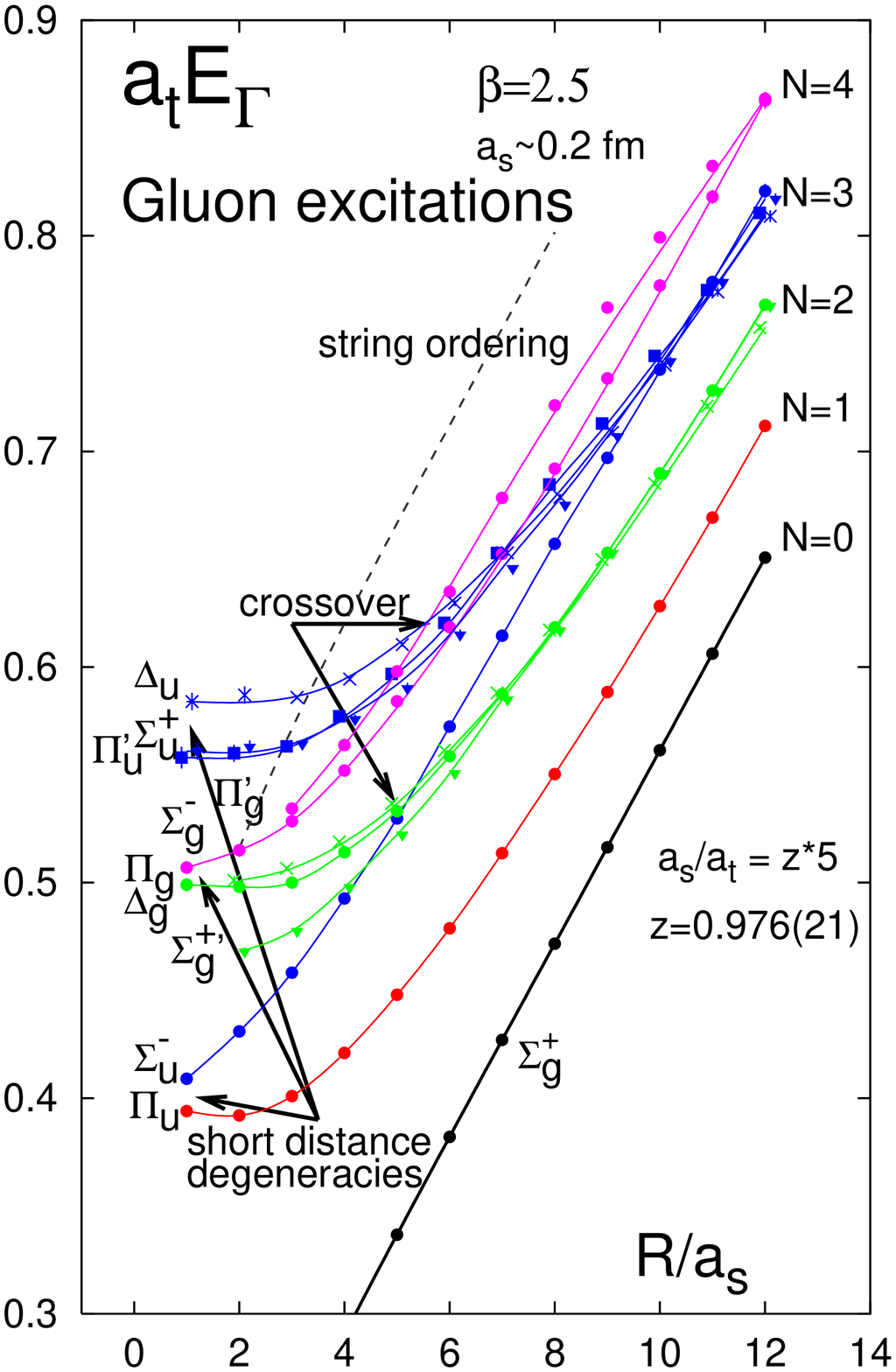}}
\caption{Hybrid static energies, $E_\Gamma$, in lattice units, from~\cite{Juge:2002br}.\label{figEg}}
\end{figure}

The gluonic static energies, $E_\Gamma$, are classified according to representations of the symmetry group $D_{\infty\,h}$, typical of molecules, 
and labeled by $\Lambda_\eta^\sigma$ (see Fig.~\ref{figsym}):
$\Lambda$ is the rotational quantum number $|\hat{\bf n}\cdot{\bf K}| = 0,1,2,\dots$,
with ${\bf K}$ the angular momentum of the gluons,
that corresponds to $\Lambda = \Sigma, \Pi, \Delta, \dots$;
$\eta$ is the CP eigenvalue ($+1\equiv g$ (gerade) and $-1 \equiv u$ (ungerade));
$\sigma$ is the eigenvalue of reflection with respect to a plane passing through the $Q\bar{Q}$ axis.
The quantum number $\sigma$ is relevant only for $\Sigma$ states.
In general there can be more than one state for each irreducible representation of $D_{\infty\,h}$: 
higher states are denoted by primes, e.g., $\Pi_u$, $\Pi_u'$, $\Pi_u'', \dots$ 
The static energies have been computed several times on the lattice in quenched QCD~\cite{Foster:1998wu,Juge:2002br,Bali:2003jq}.
In Fig.~\ref{figEg} we show the hybrid potentials computed in~\cite{Juge:2002br}.

In the limit $r\to 0$ the group $D_{\infty\,h}$ becomes the more symmetric group O(3)$\times$C.
This means that several different $\Lambda_\eta^\sigma$ representations reduce to the same $J^{PC}$ representation in that limit.
The corresponding static energies become degenerate~\cite{Foster:1998wu,Brambilla:1999xf}.
In particular, the gluelump multiplets ($\Sigma_u^-$, $\Pi_u$), ($\Sigma_g^{+\prime}$, $\Pi_g$),  ($\Sigma_g^-$, $\Pi_g^\prime$, $\Delta_g$),
($\Sigma_u^+$, $\Pi_u^\prime$, $\Delta_u$) are degenerate in the $r\to 0$ limit.
In the following, we will consider hybrids that are vibrational modes of the lowest-lying static energies $\Pi_u$ and $\Sigma_u^-$.
In the $r \to 0$ limit $\Pi_u$ and $\Sigma_u^-$ are degenerate and correspond to a gluonic operator with quantum numbers~$1^{+-}$.

\section{Schr\"odinger equation and mixing}
Quarkonium hybrid states that are vibrational modes of the $\Pi_u$ and $\Sigma_u^-$  potentials sitting in the short-range tail
of the potentials mix because of the degeneracy of the potentials in the short range.
Modes associated to higher potentials are separated from the $\Pi_u$ and $\Sigma_u^-$ potentials by an energy gap of order $\lQ$
and are integrated out with that scale.
The energies, $\mathcal{E}_N$, of the vibrational modes are much smaller than $\lQ$ and solutions of the coupled Schr\"odinger equations~\cite{Berwein:2015vca}
\bea
  &&\hspace{-1cm}
  \left[-\frac{1}{Mr^2}\,\partial_rr^2\partial_r+\frac{1}{Mr^2}
  \begin{pmatrix} l(l+1)+2 & 2\sqrt{l(l+1)} \\
    2\sqrt{l(l+1)} & l(l+1)
  \end{pmatrix}
  +
  \begin{pmatrix} E_{\Sigma_u^-}    & 0 \\
    0 & E_{\Pi_u}
  \end{pmatrix}\right]\hspace{-4pt}
\begin{pmatrix} \psi_\Sigma^{(N)} \\
  \psi_{-\Pi}^{(N)}
\end{pmatrix}  
= \mathcal{E}_N
\begin{pmatrix} \psi_\Sigma^{(N)} \\
  \psi_{-\Pi}^{(N)}
\end{pmatrix}\,,
\label{eq1}\\
&&\hspace{-1cm}
\left[-\frac{1}{Mr^2}\,\partial_r\,r^2\,\partial_r+\frac{l(l+1)}{Mr^2}+ E_{\Pi_u} \right]\psi_{+\Pi}^{(N)}=\mathcal{E}_N\,\psi_{+\Pi}^{(N)}\,,
\label{eq2}
\eea
where $l$ is the orbital angular momentum quantum number and $r$ the heavy quark-antiquark distance.
These equations follow from the equations of motion of the hybrid fields in pNRQCD.
The extension of these equations to include spin interactions is under preparation~\cite{BLSTV:2017}.
The functions  $\psi_\Sigma^{(N)}$ and $\psi_{\pm \Pi}^{(N)}$ are radial wave functions;
$\psi_\Sigma^{(N)}$ and $\psi_{-\Pi}^{(N)}$ have negative parity and $\psi_{+\Pi}^{(N)}$ positive one.
The off-diagonal terms change the $\Sigma$ wave function to $\Pi$ and vice versa, but they do not change the parity.
Hence $\psi_\Sigma^{(N)}$ mixes only with $\psi_{-\Pi}^{(N)}$, and $\psi_{+\Pi}^{(N)}$ decouples.
For $l=0$ the off-diagonal terms vanish, so the equations for $\psi_\Sigma^{(N)}$ and $\psi_{-\Pi}^{(N)}$ decouple;
there exists only one parity state, and its radial wave function is given by a Schr\"odinger equation with the $E_\Sigma$ potential and an angular part $2/Mr^2$.
The eigenstates of the equations \eqref{eq1} and \eqref{eq2} are organized in the multiplets shown in Table~\ref{tabmultiplets}.

\begin{table}[ht]
\centerline{
 \begin{tabular}{|c|c|c|c|c|}
 \hline
        & $\,l\,$ & $J^{PC}\{s=0,s=1\}$       & $E_\Gamma$           \\  \hline
  $H_1$ & $1$     & $\{1^{--},(0,1,2)^{-+}\}$ & $E_{\Sigma_u^-}$, $E_{\Pi_u}$ \\
  $H_2$ & $1$     & $\{1^{++},(0,1,2)^{+-}\}$ & $E_{\Pi_u}$               \\
  $H_3$ & $0$     & $\{0^{++},1^{+-}\}$       & $E_{\Sigma_u^-}$          \\
  $H_4$ & $2$     & $\{2^{++},(1,2,3)^{+-}\}$ & $E_{\Sigma_u^-}$, $E_{\Pi_u}$ \\
  $H_5$ & $2$     & $\{2^{--},(1,2,3)^{-+}\}$ & $E_{\Pi_u}$               \\
  \hline
 \end{tabular}}
\caption{$J^{PC}$ multiplets with $l\leq2$ for the $\Sigma_u^-$ and $\Pi_u$ gluonic states. 
  The last column shows the gluonic static energies that appear in the Schr\"odinger equation of the respective multiplet.
\label{tabmultiplets}}
\end{table}

Keeping in the equations \eqref{eq1} and \eqref{eq2} only the heavy quark-antiquark kinetic energy and the hybrid static energies, $E_\Gamma$,
amounts at the {\em Born--Oppenheimer approximation}.
Keeping only the diagonal terms amounts at the {\em adiabatic approximation}.
The exact leading order equations include both diagonal and off-diagonal terms that define the so-called {\em non-adiabatic coupling}.
As it is clear from \eqref{eq1} and \eqref{eq2} both diagonal and off-diagonal terms contribute at the same order to the energy levels in the quarkonium hybrid case.
This situation is different from the case of (electromagnetic) molecules, where the non-adiabatic coupling is subleading~\cite{Brambilla:2017uyf}.

\begin{figure}[ht]
\makebox[3.5cm]{\phantom b}\put(0,0){\epsfxsize=9truecm \epsffile{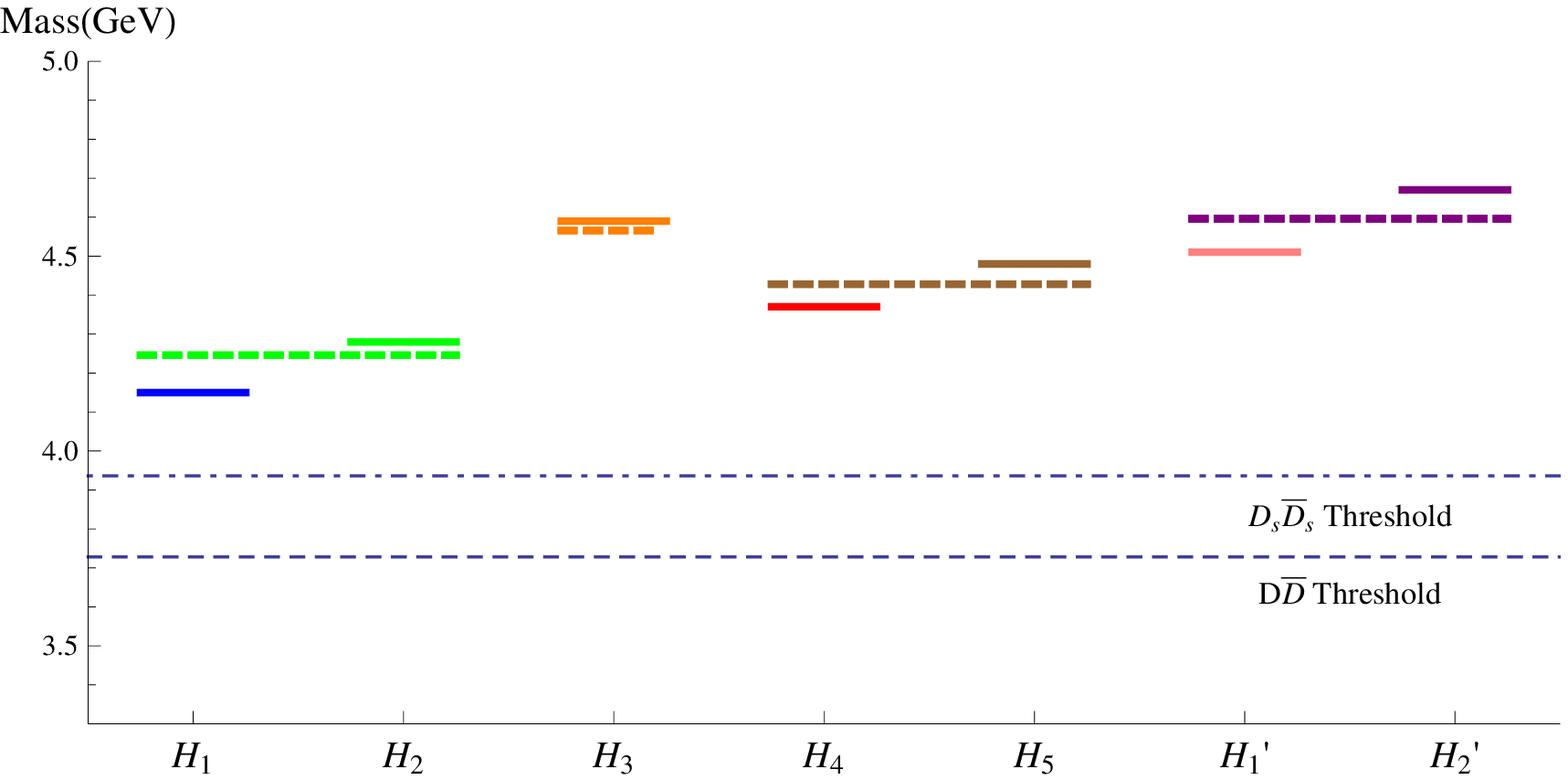}}\put(30,20){charmonium hybrids}\\
\makebox[3.5cm]{\phantom b}\put(0,-90){\epsfxsize=9truecm \epsffile{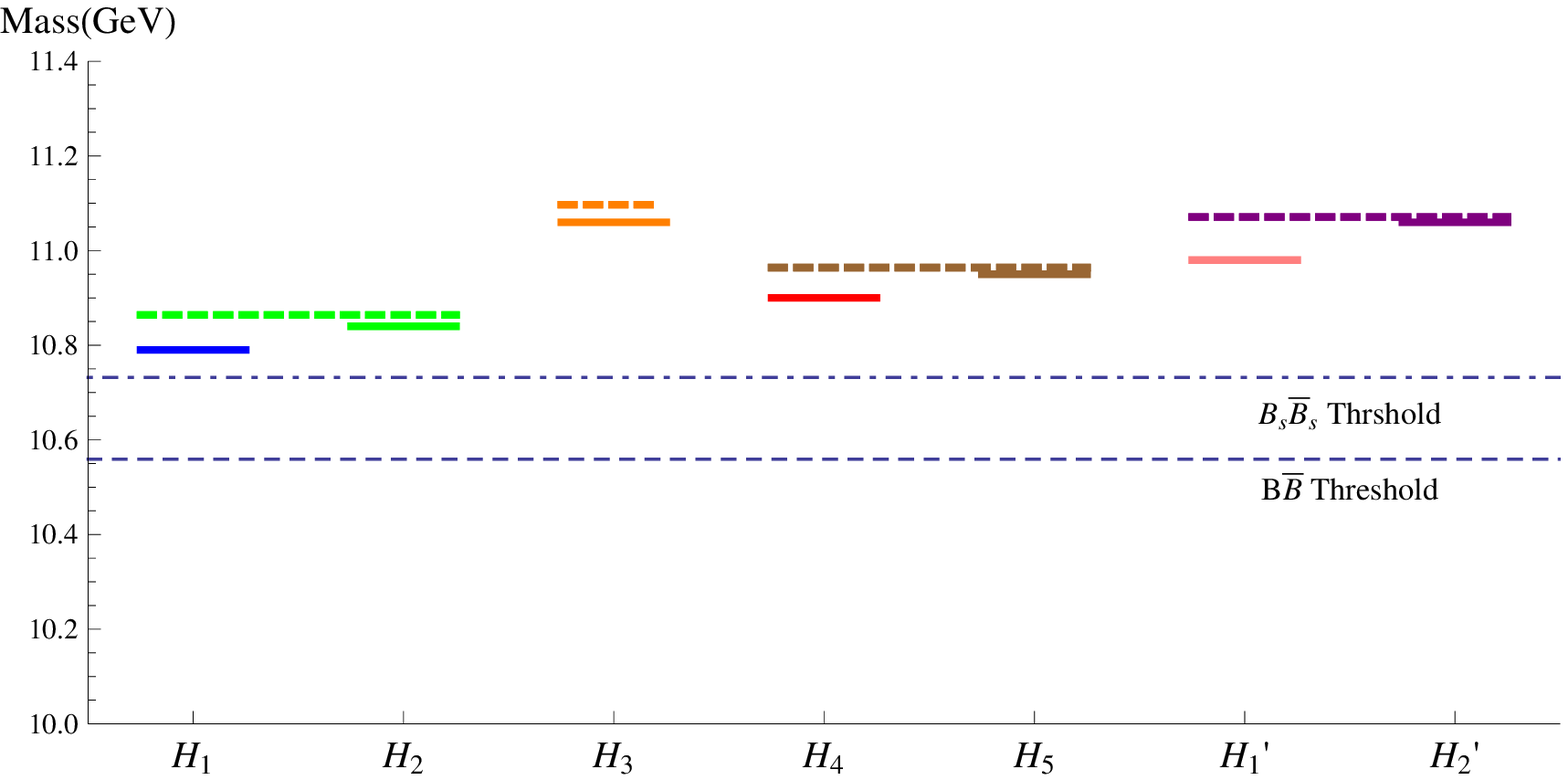}}\put(30,-70){bottomonium hybrids}\\
\caption{Comparison of the hybrid multiplet masses in the charmonium
  (upper figure) and bottomonium (lower figure) spectra obtained without mixing in~\cite{Braaten:2014qka,Braaten:2014ita,Braaten:2013boa}
  before adjusting to lattice data (dashed lines) with the results of~\cite{Berwein:2015vca} (continuous lines).
  \label{figlambdadoubling}}
\end{figure}

\section{$\Lambda$ doubling}
A physical consequence of the mixing is the so-called {\it $\Lambda$ doubling}, i.e., the lifting of degeneracy between
states with the same parity. We show this effect in Fig.~\ref{figlambdadoubling}.
The effect is also present in molecular physics, however, $\Lambda$ doubling is a subleading effect there
while it is a leading order effect in the spectrum of quarkonium hybrids.

\begin{figure}[ht]
\makebox[3.5cm]{\phantom b}\put(0,0){\epsfxsize=9truecm \epsffile{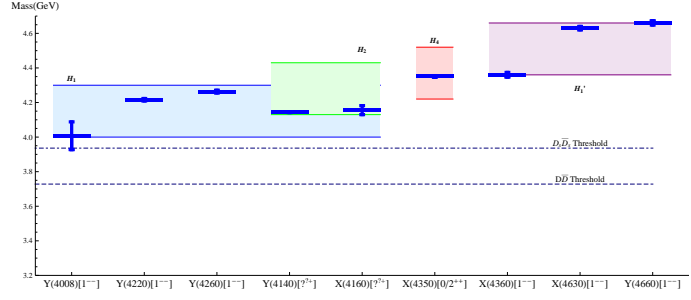}}
\caption{Comparison of some experimental candidate masses in the charmonium sector with the results from Eqs. \eqref{eq1} and \eqref{eq2} taken from~\cite{Berwein:2015vca}.
    \label{figexp}}
\end{figure}

\section{Quarkonium hybrid states versus experiments}
Several charmonium-like states have been found by the B factories in the last decade (for a recent review see~\cite{Brambilla:2014jmp}).
We compare some of these states with the hybrid spectrum obtained from solving the coupled Schr\"odinger equations \eqref{eq1} and \eqref{eq2} in Fig.~\ref{figexp}.
In the bottomonium sector, the state $Y_b(10890)[1^{--}]$ with mass $M_{Y_b} = (10.8884 \pm 3.0)$~GeV found by BELLE may be a possible $H_1$ candidate,
for which we find $M_{H_1} = (10.79 \pm 0.15)$~GeV.

\begin{figure}[ht]
\makebox[3.5cm]{\phantom b}\put(0,0){\epsfxsize=9truecm \epsffile{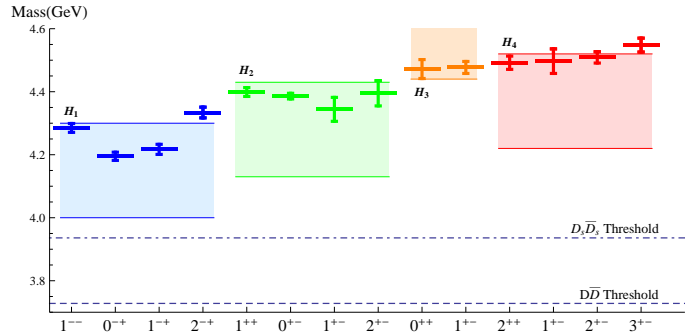}}
\caption{Comparison of results from the direct lattice computations of charmonium hybrid masses of~\cite{Liu:2012ze} with the results of~\cite{Berwein:2015vca}.
\label{figlatc}}
\end{figure}

\section{Quarkonium hybrid states versus direct lattice data}
\begin{figure}[ht]
\makebox[3.5cm]{\phantom b}\put(0,0){\epsfxsize=9truecm \epsffile{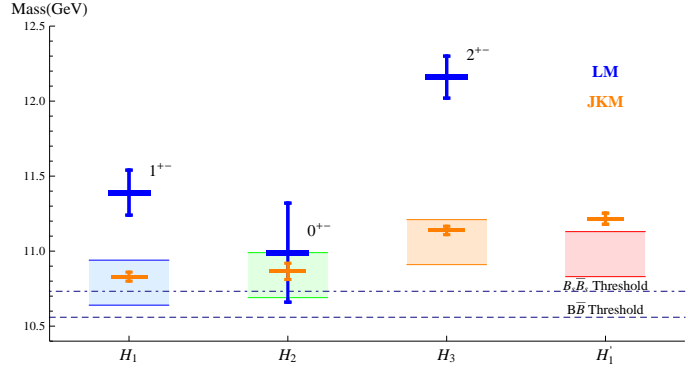}}
\caption{Comparison of results from the direct lattice computations of bottomonium hybrid masses of~\cite{Juge:1999ie} (JKM)
and~\cite{Liao:2001yh} (LM) with the results of~\cite{Berwein:2015vca}.
\label{figlatb}}
\end{figure}

In Fig.~\ref{figlatc} we compare the lattice data from~\cite{Liu:2012ze} for charmonium hybrid states 
with the spectrum from \eqref{eq1} and \eqref{eq2}.
The analogous comparison for bottomonium hybrid states of direct lattice data from~\cite{Juge:1999ie} and~\cite{Liao:2001yh}
with the spectrum from \eqref{eq1} and \eqref{eq2} is in Fig.~\ref{figlatb}.
We observe that the lattice data for charmonium hybrids are sensitive to the $\Lambda$ doubling.

\section{Conclusions}
Hybrids are one of the most specific exotics expected from QCD. 
Their experimental hunt and theoretical study has a history characterized for long time by few and difficult progresses. 
This may change with heavy quarkonium hybrids. The reason is twofold.

First, the quarkonium spectrum above threshold has been extensively investigated by experiments in the last decade 
providing an enormous wealth of new and unexpected states and many new data. Some of these are undoubtedly exotic.
Some may be indeed hybrids made of a heavy quark-antiquark pair.

Second, effective field theories for heavy quarkonia combined with lattice QCD offer a systematic
and rigorous tool (based on the nonrelativistic hierarchy of energy scales) to study many of the quarkonium hybrid observables from QCD: spectra, splittings, decays.
A crucial theoretical progress will come from improved lattice determinations of the hybrid potentials.
So far no unquenched lattice simulation is available for them.
Also systematic calculations of higher-order potentials, which are necessary to understand spin splittings, are missing.

The effective field theory framework developed to describe quarkonium hybrids may be used to describe also tetraquarks.
In this case, it is necessary to compute on the lattice the heavy-quark potentials with a light-quark pair
playing a similar role to the gluonic excitation in the hybrid case.
Such potentials, which are poorly or not known, have been identified in the last couple of years as a crucial ingredient
to provide for the first time a dynamical description of tetraquarks in QCD.
Also mixing of tetraquarks with other quarkonium states could be studied and computed in this framework.

\end{document}